\documentclass[sigconf]{acmart}

\usepackage{cancel}
\usepackage{algorithm}
\usepackage{algpseudocode}
\usepackage{geometry}
\usepackage{subcaption}  
\usepackage{bm}
\usepackage{wrapfig}
\usepackage{graphicx}

\AtBeginDocument{}

\copyrightyear{2025} 
\acmYear{2025} 
\setcopyright{acmlicensed}\acmConference[SA Technical Communications '25]{SIGGRAPH Asia 2025 Technical Communications}{December 15--18, 2025}{Hong Kong, Hong Kong}
\acmBooktitle{SIGGRAPH Asia 2025 Technical Communications (SA Technical Communications '25), December 15--18, 2025, Hong Kong, Hong Kong}
\acmDOI{10.1145/3757376.3771398}
\acmISBN{979-8-4007-2136-6/2025/12}

\acmSubmissionID{149}

\begin{document}

\title{Modeling and Simulating Origami Structures using Bilinear Solid-Shell Element}


\author{Qixin Liang}
\orcid{0009-0002-5494-3672}
\affiliation{%
  \institution{The University of Hong Kong}
  \city{Hong Kong}
  \country{Hong Kong}
}
\affiliation{%
  \institution{TransGP}
  \city{New Territories}
  \country{Hong Kong}
}
\email{liangqixin23@outlook.com}

\begin{abstract}
    We propose a novel computational framework for modeling and simulating origami structures. In this framework, bilinear solid-shell elements are employed to model the origami panels while crease folding is considered through the angle between the director vectors of the adjacent panels. The director vector is the vector normal to the mid-surface before displacement/deformation comes in. To mitigate locking issues in the solid-shell element, we introduce the assumed natural strain method. To validate the effectiveness of our framework, we conduct origami simulations involving both straight- and curved-creases. The accuracy and efficacy of the framework are demonstrated through quantitative and qualitative analyses.
\end{abstract}



\begin{teaserfigure}
  \includegraphics[width=0.95\textwidth]{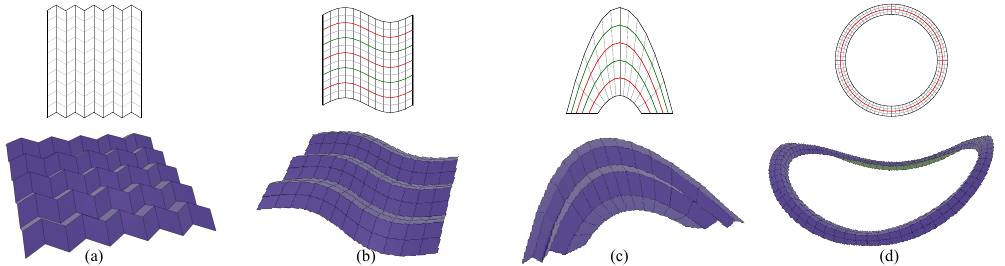}
  \caption{Our framework enables the modeling and simulation of origami structures with pre-creased straight and curved fold lines using bilinear solid-shell elements, as demonstrated in these images. This approach provides an effective framework for the design and analysis of origami structures.}
  \label{fig:teaser}
  \Description{Our framework enables the modeling and simulation of origami structures with pre-creased straight and curved fold lines using bilinear solid-shell elements. Our approach provides an effective framework for the design and analysis of origami structures.}
\end{teaserfigure}


\maketitle

\section{Introduction}

Origami, rooted in the ancient art of paper folding, has evolved into a multidisciplinary field of science and engineering. As origami structures have transitioned from rigid panels to deformable panels exhibiting complex multi-physics responses, there is a growing need for simulation techniques that accurately capture both the geometric and physical behaviors.

Origami simulation methods can be broadly classified into kine-matics-based and mechanics-based ones. 
\textit{Kinematics-based} methods~\cite{tachi2009simulation} assume panels remain rigid and planar, enabling folding angles only as variables to describe deformation. These methods are computationally efficient but cannot consider panel deformations. 
\textit{Mechanics-based} methods relax the rigidity assumption, allowing membrane and bending deformations. 
The widely used bar-hinge model~\cite{liu2017nonlinear}, similar to the mass–spring model in cloth simulation~\cite{Bridson2005}, suffers from mesh-dependent issues, with material parameters that vary with mesh resolution and are not easily transferable across different mesh topologies.
Discrete shell models~\cite{Burgoon2006DiscreteSO} improve membrane accuracy of the origami panels but still exhibit mesh-dependent bending behavior. 
Other discrete and ruling-based models~\cite{Rabinovich2019, Solomon2012} impose strict isometry constraints, making them difficult to consider physical constitutive laws.
In~\cite{hu2021simulating}, a corotational quadrilateral element was proposed to model the bending deformation of quadrilateral panels; however, it is limited to capturing only the warping bending within individual elements.

Motivated by these advancements, we introduce a solid-shell element to model origami panels. Instead of using discrete rods to represent deformed fold lines~\cite{Le2023}, relying on remeshing to smooth sharp creases~\cite{Narain2013}, or employing the normals of rigid prismatic panels~\cite{Kilian2017}, we preserve the sharp features and conveniently model folding motion through the angle between director vectors of the elements along the same crease. Our contributions are:

\begin{itemize}
\item We propose a crease model based on the director vector of the solid-shell element.
\item We present a total Lagrangian formulation for a bilinear quadrilateral solid-shell element~\cite{Sze2002IJNME}, which effectively mitigates various locking phenomena through the assumed natural strain (ANS) method.
\item We present two computational origami folding examples—one with a straight crease and the other with a curved crease—along with their analytical solutions, boundary conditions, and material setups. They may serve as useful references for benchmarking origami simulation tools.
\end{itemize}

This work aims to provide a versatile tool for the design and analysis of origami structures.  


\section{Computational model for origami simulation}

\subsection{Solid-Shell Finite Element Model}

\begin{figure} [h] 
  \centering
  \includegraphics[width=0.65\linewidth]{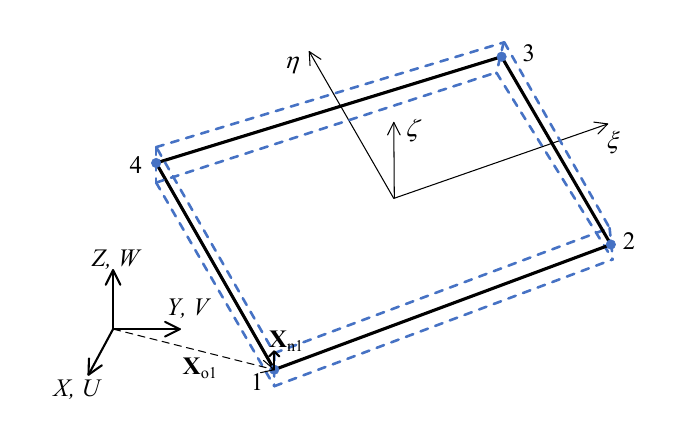} \label{fig:Q4}
  \Description{Parameterization of a bilinear solid-shell element using $\xi$, $\eta$, and $\zeta$. 
   The black solid quadrilateral represents the mid-surface of the element, parameterized by $\xi$ and $\eta$, while $\zeta$ parameterizes the through-thickness direction.}
\end{figure}

The above figure shows a bilinear solid-shell element with nodes on its mid-surface, the interpolated initial position vector $\mathbf{X}$ and displacement vector $\mathbf{U}$ can be written as:
\begin{equation}
\begin{gathered}
    \mathbf{X} = \mathbf{X}_{o}(\xi, \eta) + \zeta \mathbf{X}_{n}(\xi, \eta) = \sum_i N_i(\xi, \eta)\, \mathbf{X}_{oi} + \zeta \sum_i N_i(\xi, \eta)\, \mathbf{X}_{ni}, \\
    \mathbf{U} = \mathbf{U}_{o}(\xi, \eta) + \zeta \mathbf{U}_{n}(\xi, \eta) = \sum_i N_i(\xi, \eta)\, \mathbf{U}_{oi} + \zeta \sum_i N_i(\xi, \eta)\, \mathbf{U}_{ni},
\end{gathered}
\end{equation}
where $(\xi, \eta, \zeta)$ is the natural coordinate vector $\boldsymbol{\xi}$ and and its components are bounded by -1 and 1; $N_{i}$s are standard bilinear interpolation functions; 
$\mathbf{X}_{o} = [X_o, Y_o, Z_o]^{T}$ and $\mathbf{X}_{n}= [X_n, Y_n, Z_n]^{T}$ are the initial midsurface position and director vectors, while $\mathbf{U}_{o}= [U_o, V_o, W_o]^{T}$ and $\mathbf{U}_{n}= [U_n, V_n, W_n]^{T}$ are the midsurface and director displacement vectors.

The initial nodal director $\mathbf{X}_{ni}$ should be taken to be perpendicular to the actual initial mid-surface which, unless is flat, is different from the interpolated mid-surface $\mathbf{X}_o(\xi, \eta)$. The natural Green-Lagragian strain tensor is
\begin{equation}
    \boldsymbol{\varepsilon} = \frac{1}{2}\left( \left( \frac{\partial (\mathbf{X} + \mathbf{U})}{\partial \boldsymbol{\xi}} \right)^{T} \frac{\partial (\mathbf{X} + \mathbf{U})}{\partial \boldsymbol{\xi}} 
    - \left( \frac{\partial \mathbf{X} }{\partial \boldsymbol{\xi}} \right)^{T} \frac{\partial \mathbf{X} }{\partial \boldsymbol{\xi}} \right).
\end{equation}
Its components include $\varepsilon_{\phi\psi}$, $\gamma_{\zeta\phi}$ ($\phi,\psi = \xi, \eta$), and $\varepsilon_{\zeta\zeta}$, corresponding to the natural inplane, transverse shear, and transverse normal strains, respectively. More details about the element formulation and the assumed natural strain method are provided in the supplement material.

\subsection{Crease Modeling}


\begin{figure}[h]
  \centering
  \includegraphics[width=0.5\linewidth]{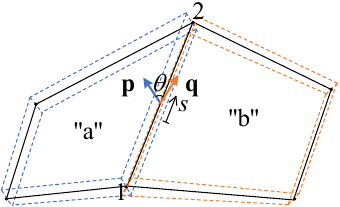}
  \Description{Two origami panels are connected by a common crease, while the directors of the two elements along the crease are independent.}
  \label{fig:Q4_angle}
\end{figure}

In the above Figure, $s \in [-1, 1]$ is the parametric coordinate of the crease between elements "a" and "b". While the crease is common, the directors of the two elements along the crease are independent. In the current configuration, let $\mathbf{p}$ and $\mathbf{q}$ denote the director of elements "a" and "b", respectively. They can be obtained by interpolating the deformed director $\mathbf{X}_{n1}+\mathbf{U}_{n1}$ at node 1, $s=-1$ and $\mathbf{X}_{n2}+\mathbf{U}_{n2}$ at node 2, $s=1$, as
\begin{equation} 
\begin{aligned}
&\mathbf{p} = L_{1}(\mathbf{X}_{n1}^{a}+\mathbf{U}_{n1}^{a}) + L_{2}(\mathbf{X}_{n2}^{a}+\mathbf{U}_{n2}^{a}), \\
&\mathbf{q} = L_{1}(\mathbf{X}_{n1}^{b}+\mathbf{U}_{n1}^{b}) + L_{2}(\mathbf{X}_{n2}^{b}+\mathbf{U}_{n2}^{b}), 
\end{aligned}
\end{equation}
where $L_1 = (1-s)/2$ and $L_2 = (1+s)/2$ are linear interpolation functions.

In thin shell analyses, transverse shear and thickness strain in the solid-shell strain elements are enforced to zero, the through-thickness direction is effectively inextensible, and the tangential relative motion of directors $\mathbf{p}$ and $\mathbf{q}$ along the crease is negligible. Therefore, the folding angle of crease can be approximated by
\begin{equation} 
\theta= 
\begin{cases}
\cos ^{-1} \frac{\mathbf{p} \cdot \mathbf{q}}{p q} \quad \text{for} \ (\mathbf{p} \times \mathbf{q}) \cdot(\mathbf{x}_{o2} - \mathbf{x}_{o1}) \geq 0 \\ 
-\cos ^{-1} \frac{\mathbf{p} \cdot \mathbf{q}}{p q} \quad \text{otherwise}
\end{cases}.
\end{equation}
In the case that the two elements are flat, $\theta$ ranges from $[-\pi, \pi]$ with $\theta = \pm \pi$ indicating that the elements are fully folded. To prevent the physically inadmissible self-intersection configuration, the following crease energy $\Psi_{c}$ with nonlinear folding constitutive model is adopted
\begin{equation}
    l \left\{
    \begin{array}{l@{\quad}r}  
        \begin{aligned}
            &\frac{1}{2}k_f (\theta_0 - \theta_L)^2 + k_f (\theta_0 - \theta_L)(\theta_L - \theta) \\
            &\quad - \frac{4k_f (\theta_L + \pi)^2}{\pi^2} \ln\left|\cos\left( \frac{\pi(\theta_L - \theta)}{2(\theta_L + \pi)} \right)\right|
        \end{aligned} & -\pi < \theta < \theta_L \\[2ex]
        \begin{aligned}
        \frac{1}{2}k_f (\theta - \theta_0)^2
        \end{aligned} & \theta_L \leq \theta \leq \theta_R \\[2ex]
        \begin{aligned}
            &\frac{1}{2}k_f (\theta_R - \theta_0)^2 + k_f (\theta_R - \theta_0)(\theta - \theta_R) \\
            &\quad - \frac{4k_f(\pi - \theta_R)^2}{\pi^2} \ln\left|\cos\left( \frac{\pi (\theta - \theta_R)}{2(\pi - \theta_R)} \right)\right|
        \end{aligned} & \theta_R < \theta < \pi
    \end{array}
    \right. ,
\end{equation}
where $l = \Vert \mathbf{X}_{o2} - \mathbf{X}_{o1} \Vert$ is the crease length, and $k_f$ is the folding stiffness per unit length~\cite{liu2017nonlinear}. 
The values $\theta_L$ and $\theta_R$ are fold angle limits that activate the penalty. 
Within the interval $[\theta_L, \theta_R]$, the energy exhibits a standard quadratic behavior centered at the rest angle $\theta_0$. Outside this range, the logarithmic term grows rapidly to impose strong penalty on configurations approaching $\theta = \pm \pi$, thereby effectively preventing self-intersection.

\section{Results}
Our computational model can be implemented in most, if not all, finite element programs. Here, we use a damped Newton solver with the adaptive increment method  (see our supplemental document). In the tests below, SI units are used in expressing all material and geometric parameters. 
The quantitative validation includes two benchmark tests with known analytical solutions and an additional case presented in the supplemental document to further verify the solid-shell element. Subsequently, a series of qualitative tests is provided to illustrate the applicability of the present framework to more complex scenarios.
\subsection{Quantitative tests}


\begin{figure}
  \centering
  \includegraphics[width=0.99\linewidth]{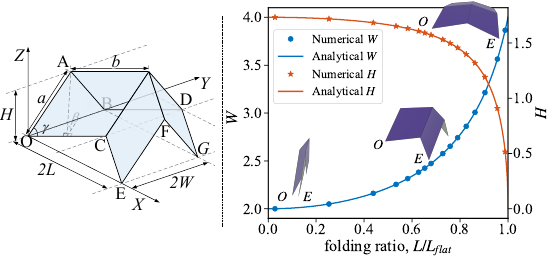}
  \caption{\textit{Compressing the Miura-ori unit cell.} Left: A Miura-ori unit cell. Right: The width $W$ and height $H$ plotted against the folding ratio $L / L_{{flat}}$. 
  }
  \label{fig:miura_unit_curve}
  \Description{\textit{Compressing the Miura-ori unit cell.} The width $W$ and height $H$ plotted against the folding ratio $L / L_{{flat}}$}
\end{figure}



\subsubsection{Compressing the Miura-ori unit cell.} 

A Miura-ori unit cell is illustrated in the left image of Figure~\ref{fig:miura_unit_curve} and a complete Miura-ori structure is shown in Figure~\ref{fig:teaser}a. During the folding process, the cell deforms solely along the predefined creases. As a result, its deformation can be described analytically under the rigid panel assumption~\cite{Schenk2013}. The analytical solution is given by
\begin{equation} 
\begin{gathered} H = a \sin \beta \sin \gamma, \\ \ L=2b\frac{\cos \gamma \tan\beta }{\sqrt{1+\cos^{2}\gamma \tan^2\beta}}, \quad W=2a\sqrt{1-\sin^2\gamma \sin^2\beta}.
\end{gathered} 
\end{equation}
The geometry of the Miura-ori unit cell is defined by the parameters $a = 2$, $b = 2$, and $\gamma = 60^\circ$. 
To model the Miura-ori unit cell as a "rigid origami" system—consisting of rigid panels connected by compliant hinges—we follow the approach in~\cite{liu2017nonlinear}, setting the bending rigidity of the panels to be $10^5$ times greater than the folding stiffness. The material parameters used are $E = 12 \times 10^9$, $\nu = 0.3$, $h = 0.01$, and $k_f = 0.01$. 
To avoid buckling, $\beta$ is set to be $15^\circ$ in the initial configuration, making the Miura-ori unit cell nearly flat. The boundary conditions are imposed as follows: $\mathbf{U}_o = \mathbf{0}$ at node O; the $U_o=0$ at nodes A and B; the $W_o=0$ at nodes O, B, C, D, E, and G; $U_o = -3.44$ is prescribed to nodes E, F, and G to compress the unit cell into a folded configuration.
As shown in the right image of Figure~\ref{fig:miura_unit_curve}, our predictions are indistinguishable from the analytical solution.

\begin{figure}
  \centering
  \includegraphics[width=0.85\linewidth]{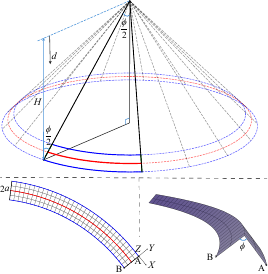}
  \caption{\textit{Folding a creased annulus sector into a theoretical cone.} The $32 \times 4$ meshed cut annulus shown at the bottom left is extracted from the cone at the top, which has an apex angle of $\phi = 90^\circ$. The curved-crease origami structure on the bottom right also exhibits a fold angle of $\phi = 90^\circ$. 
  }
  \label{fig:annulus_sector}
  \Description{Folding a creased annulus sector into a theoretical cone.}
\end{figure}

\begin{figure}
  \centering
  \includegraphics[width=0.99\linewidth]{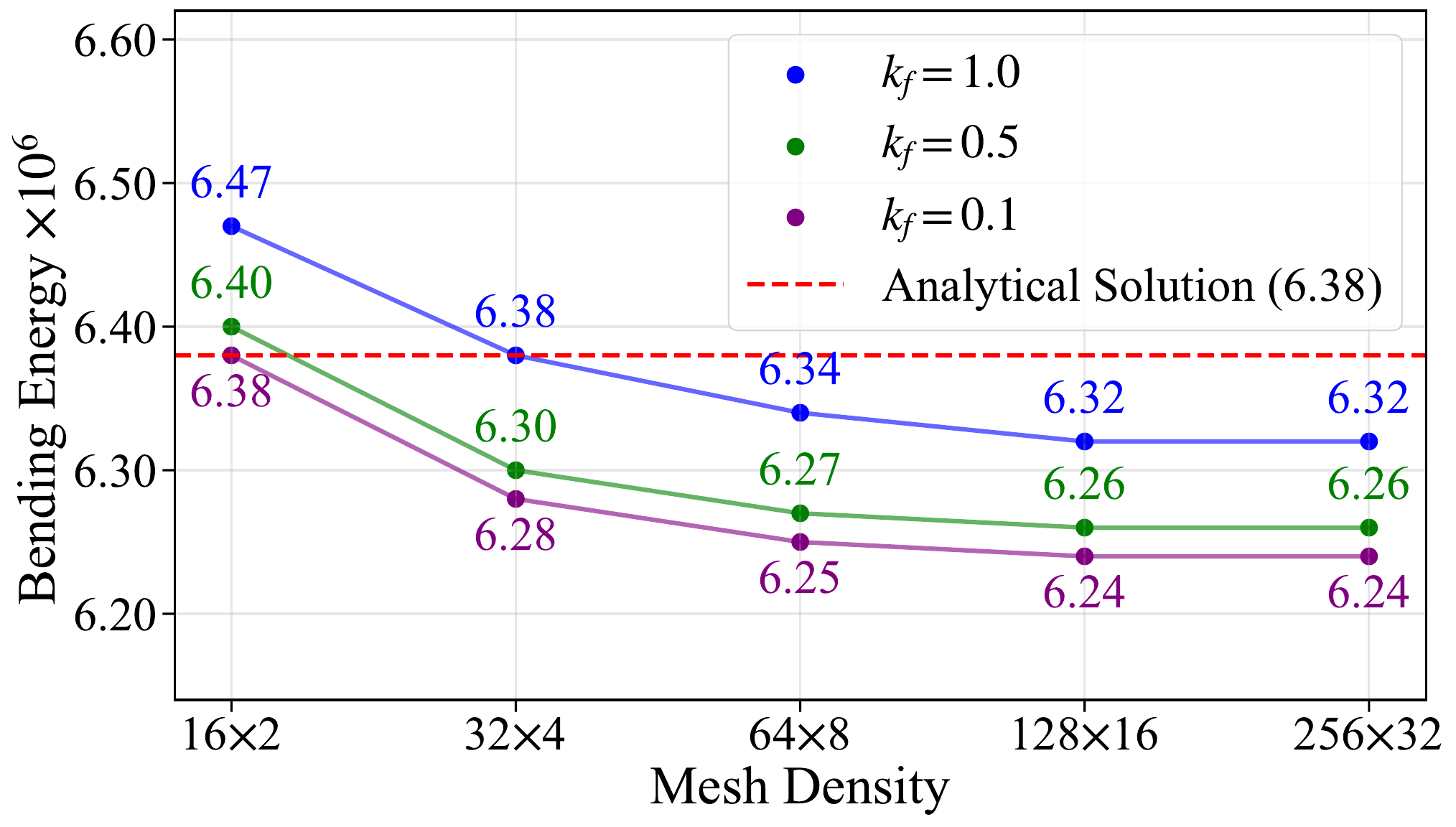}
  \caption{The bending energy in the cut annulus versus the mesh density. 
  }
  \label{fig:mesh_convergence}
  \Description{The bending energy in the cut annulus versus the mesh density.}
\end{figure}
\subsubsection{Folding a creased annulus sector into a theoretical cone.}

Following~\cite{Woodruff2020}, we cut an annulus sector from a cone, flatten it into a planar sheet, and then fold it along the curved crease into a curved-crease structure (see Figure~\ref{fig:annulus_sector}). Folding the initially flat sheet along a curved crease involves bending the developable cone surfaces during the motion. Under the isometric assumption, the theoretical bending energy of the annular region on the right cone surface can be computed as  
\begin{equation}
    \Psi_{\text{annulus}} = \frac{1}{2} \int_{\Omega} D_b \kappa_{p}^2 \, d\Omega,
\end{equation}  
where \( D_b = {E h^3}/{12(1 - \nu^2)} \) is the bending rigidity of an isotropic shell, $\Omega$ is the area of the annular region, and the principal curvature $\kappa_p$ is 
\begin{equation}
    \kappa_p = \frac{1}{\tan(\phi/2) \,  \sqrt{1 + \tan^2(\phi/2)}d} ,
\end{equation}  
where $d$ is the distance from the apex of the cone measured along its height, and $\phi/2$ is half of the cone’s apex angle (see Figure~\ref{fig:annulus_sector}). Following~\cite{Woodruff2020}, when the fold angle of the curved-crease structure is $\phi=90^\circ$, the resulting folded shape becomes an exact segment of a cone.

In the numerical simulation, we adopt the geometric and material parameters from~\cite{Woodruff2020}. The radius of the middle arc is $R = 0.1$, which bisects the annulus of width $a = 0.005$, and the central angle is $\alpha = \pi/4$. $E = 4 \times 10^{9}$, $\nu = 0$, and $h = 0.1 \times 10^{-3}$. The folding stiffness $k_f$ is typically scaled relative to the bending rigidity of the shell~\cite{Lechenault2014}; hence, we consider three representative values: 0.1, 0.5, and 1.
The boundary conditions are imposed as follows: $\mathbf{U}_{o}=\mathbf{0}$ at node A; $V_{o}=W_{o}=0$ at node B; $W_o=0$ is also prescribed for all nodes along the inner and outer arcs; $W_o = a/\sqrt{2}$ is applied to the middle arc to induce a folding angle of $90^\circ$.

As shown in Figure~\ref{fig:mesh_convergence}, the predicted bending energy rapidly converges close to the theoretical value as the mesh is refined, with relative errors of approximately $0.9\%$, $1.8\%$, and $2.1\%$ corresponding to folding stiffness values of 1, 0.5, and 0.1, respectively. This clearly illustrates the influence of the folding stiffness on the bending response. The case highlights the potential of our method for simulating and analyzing the curved-crease origami structures.

\subsection{Qualitative tests}
\begin{figure}[t]
  \centering
  \includegraphics[width=0.78\linewidth]{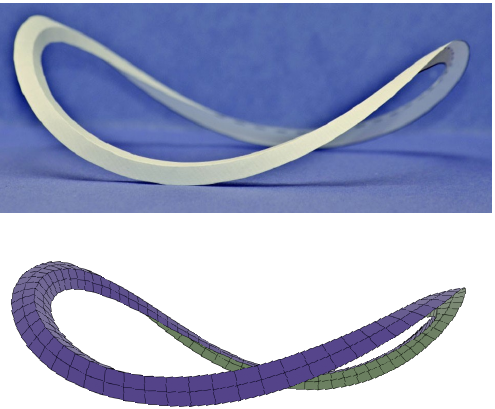}
  \caption{\textit{Buckle the paper-made full annulus.} 
  The top image is reprinted with permission from 
  Marcelo A. Dias, Levi H. Dudte, L. Mahadevan, and Christian D. Santangelo, 
  \textit{Phys. Rev. Lett.} 109, 114301 (2012). 
  Copyright \textcopyright 2012 American American Physical Society. (DOI: \url{http://dx.doi.org/10.1103/PhysRevLett.109.114301})
  The below one is our simulated result.}
  \label{fig:full_annulus}
  \Description{Comparison of a paper-made annulus buckling experiment (top) and our simulated result (bottom).}
\end{figure}

Figures~\ref{fig:teaser} and~\ref{fig:full_annulus} present additional qualitative results, illustrating our model’s capability to handle more complex structures.

\section{Conclusion and Discussion}
In this paper, we propose the use of solid-shell elements to model origami structures, where the naturally embedded director vectors of the elements are utilized to compute the crease folding angle. Origami structures with straight- and curved-creases are considered. Our predictions are close to the analytical solutions. 

Our modeling approach can be naturally extended to triangular and higher-order solid-shell elements, while still utilizing director vectors to model creases, thereby enabling the simulation of curved-tile origami structures~\cite{Liu2024}, which are challenging to capture with existing methods.

\begin{acks}
The author thanks Professor K.Y. Sze and the anonymous reviewers for their valuable comments, and gratefully acknowledges the funding supporter, TransGP. 
\end{acks}

\bibliographystyle{ACM-Reference-Format}
\bibliography{references}

\appendix

\section{Bilinear Solid-Shell Element} 

The interpolated initial position vector $\mathbf{X}$ and displacement vector $\mathbf{U}$ can be written as:
\begin{equation}
\begin{aligned}
    &\mathbf{X}=\mathbf{X}_{o}\left(\xi,\eta\right)+\zeta \mathbf{X}_{n}\left(\xi,\eta\right)=\sum_{i=1}^{4} N_{i}\mathbf{X}_{oi}+\zeta\sum_{i=1}^{4} N_{i}\mathbf{X}_{oi},\\
    &\mathbf{U}=\mathbf{U}_{o}\left(\xi,\eta\right)+\zeta \mathbf{U}_{n}\left(\xi,\eta\right)=\sum_{i=1}^{4} N_{i}\mathbf{U}_{oi}+\zeta\sum_{i=1}^{4} N_{i}\mathbf{U}_{ni},
\end{aligned}
\end{equation}
where $(\xi,\eta,\zeta)$ is the natural coordinate vector $\boldsymbol{\xi}$ and its components are bounded by -1 and 1; the interpolation functions for the $i$-th element nodes are
\begin{equation}
\begin{gathered}
    N_{1}=\frac{\left(1-\xi\right)(1-\eta)}{4}, \, N_{2}=\frac{\left(1+\xi\right)(1-\eta)}{4}, \\
    N_{3}=\frac{\left(1+\xi\right)(1+\eta)}{4}, \, N_{4}=\frac{\left(1-\xi\right)(1+\eta)}{4}.
\end{gathered}
\end{equation}
The initial nodal director $\mathbf{X}_{ni}$ should be taken to be perpendicular to the actual initial mid-surface which, unless is flat, is different from the interpolated mid-surface $\mathbf{X}_o(\xi, \eta)$. The natural Green-Lagragian strain tensor is
\begin{equation}
\begin{aligned}
    \varepsilon
&=\frac{1}{2}\left(\left(\frac{\partial\left(\mathbf{X}+\mathbf{U}\right)}{\partial\boldsymbol{\xi}}\right)^{T}\frac{\partial\left(\mathbf{X}+\mathbf{U}\right)}{\partial\boldsymbol{\xi}}-\left(\frac{\partial \mathbf{X}}{\partial\boldsymbol{\xi}}\right)^{T}\frac{\partial \mathbf{X}}{\partial\boldsymbol{\xi}}\right)\\
&=\frac{1}{2}\left(\left(\frac{\partial \mathbf{U}}{\partial\boldsymbol{\xi}}\right)^{T}\frac{\partial \mathbf{U}}{\partial\boldsymbol{\xi}}+\left(\frac{\partial \mathbf{U}}{\partial\boldsymbol{\xi}}\right)^{T}\frac{\partial \mathbf{X}}{\partial\boldsymbol{\xi}}+\left(\frac{\partial \mathbf{X}}{\partial\boldsymbol{\xi}}\right)^{T}\frac{\partial \mathbf{U}}{\partial\boldsymbol{\xi}}\right).
\end{aligned}
\end{equation}
Its components include
\begin{equation}
\begin{gathered}
    \varepsilon_{\phi\psi}=\frac{1}{2}(\mathbf{X}_{,\phi}^{T}\mathbf{U}_{,\psi}+\mathbf{X}_{,\psi}^{T}\mathbf{U}_{,\phi}+\mathbf{U}_{,\phi}^{T}\mathbf{U}_{,\psi})=\varepsilon_{m\phi\psi}+\zeta\varepsilon_{b\phi\psi}+\zeta^{2}\text{-terms},\\
\varepsilon_{\zeta\phi}=\frac{1}{2}(\mathbf{X}_{,\phi}^{T}\mathbf{U}_{n}+\mathbf{X}_{n}^{T}\mathbf{U}_{,\phi}+\mathbf{U}_{,\phi}^{T}\mathbf{U}_{n})=\frac{\gamma_{\zeta\phi}}{2}+\zeta\text{-terms}, \\
\varepsilon_{\zeta\zeta}=\mathbf{X}_{n}^{T}\mathbf{U}_{n}+\frac{\mathbf{U}_{n}^{T}\mathbf{U}_{n}}{2},
\end{gathered}
\end{equation}
where $\phi,\psi = \xi, \eta$; $\varepsilon_{m\phi\psi}$, $\varepsilon_{b\phi\psi}$, $\gamma_{\zeta\phi}$ and $\varepsilon_{\zeta\zeta}$ are the natural membrane, bending, transverse shear and transverse normal strain. The first and second order $\zeta$-terms in $\varepsilon_{\zeta\phi}$ and $\varepsilon_{\phi\psi}$ are of secondary effects and will be neglected. Further expansion gives
\begin{equation}
\begin{aligned}
    \varepsilon_{m\phi\psi}
    &=\frac{1}{2}\left(\mathbf{X}_{o,\phi}^{T}\mathbf{U}_{o,\psi}+\mathbf{X}_{o,\psi}^{T}\mathbf{U}_{o,\phi}+\mathbf{U}_{o,\phi}^{T}\mathbf{U}_{o,\psi}\right) \\
    \varepsilon_{b\phi\psi}
    &=\frac{1}{2}\left( \mathbf{X}_{o,\phi}^{T}\mathbf{U}_{n,\psi}+\mathbf{X}_{n,\phi}^{T}\mathbf{U}_{o,\psi}+\mathbf{X}_{o,\psi}^{T}\mathbf{U}_{n,\phi} \right. \\
&\quad\,\,\left. +\mathbf{X}_{n,\psi}^{T}\mathbf{U}_{o,\phi}+\mathbf{U}_{o,\phi}^{T}\mathbf{U}_{n,\psi}+\mathbf{U}_{o,\psi}^{T}\mathbf{U}_{n,\phi} \right) \\
\gamma_{\zeta\phi}
&=\mathbf{X}_{o,\phi}^{T}\mathbf{U}_{n}+\mathbf{X}_{n}^{T}\mathbf{U}_{o,\phi}+\mathbf{U}_{o,\phi}^{T}\mathbf{U}_{n}.
\end{aligned}
\end{equation}
It is well-known that solid-shell elements are prone to membrane, shear and trapezoidal lockings. The first two lockings are caused by excessive enforcements of the zero membrane and transverse shear strains. The third locking occurs when a linear/bilinear solid element is used to model a curved surface, resulting in a trapezoidal element cross-section. While membrane locking can be ignored in lower order element such as the present bilinear one, the shear and trapezoidal lockings are alleviated by ANS in which the relevant natural components are interpolated at selected boundary points. These include~\cite{Sze2002IJNME}
\begin{equation}
\begin{gathered}
    \gamma_{\zeta\xi}^{ANS}=\frac{1-\eta}{2}\left(\gamma_{\zeta\xi}\right)_{\xi=0,\eta=-1}+\frac{1+\eta}{2}\left(\gamma_{\zeta\xi}\right)_{\xi=0,\eta=+1} \\
    \gamma_{\zeta\eta}^{ANS}=\frac{1-\xi}{2}\left(\gamma_{\zeta\eta}\right)_{\xi=-1,\eta=0}+\frac{1+\xi}{2}\left(\gamma_{\zeta\eta}\right)_{\xi=+1,\eta=0} \\
    \varepsilon_{\zeta\zeta}^{ANS}=\sum_{i=1}^{4} N_{i} \cdot (\varepsilon_{\zeta\zeta})_{i},
\end{gathered}
\end{equation}
where $(\varepsilon_{\zeta\zeta})_{i}$ denotes the natural thickness strain at the i-th element node. 

With respect to a local orthogonal coordinates $(x,y,z)$ with the $x$-$y$-plane tangential to the initial mid-surface, the strain transformation relations can be expressed as:
\begin{equation}
\begin{gathered}
\boldsymbol{\varepsilon}_{m}=\begin{pmatrix}\varepsilon_{mxx}\\\varepsilon_{myy}\\2\varepsilon_{mxy}\end{pmatrix}=\mathbf{T}_{\boldsymbol{\varepsilon}}\begin{pmatrix}\varepsilon_{m\xi\xi}\\\varepsilon_{m\eta\eta}\\2\varepsilon_{m\xi\eta}\end{pmatrix}, \, \boldsymbol{\varepsilon}_{b}=\begin{pmatrix}\varepsilon_{bxx}\\\varepsilon_{byy}\\2\varepsilon_{bxy}\end{pmatrix}=\mathbf{T}_{\boldsymbol{\varepsilon}}\begin{pmatrix}\varepsilon_{b\xi\xi}\\\varepsilon_{b\eta\eta}\\2\varepsilon_{b\xi\eta}\end{pmatrix} \\
\boldsymbol{\gamma}=\begin{pmatrix}\gamma_{zx}\\\gamma_{zy}\end{pmatrix}=\mathbf{T}_{\boldsymbol{\gamma}}\begin{pmatrix}\gamma_{\zeta\xi}^{ANS}\\\gamma_{\zeta\eta}^{ANS}\end{pmatrix}, \, \varepsilon_{zz}=T_{\zeta}\varepsilon_{\zeta\zeta}^{ANS},
\end{gathered}
\end{equation}
in which $T_{\boldsymbol{\varepsilon}}$, $T_{\boldsymbol{\gamma}}$ and $T_{\zeta}$ are derived in Section~\ref{Strain_Trans}. It should be remarked that the above transformation relations are approximation only unless $\zeta$ and $z$ are parallel. The following assumptions will be made on stress-strain relationships, 
\begin{equation}
\begin{gathered}
\begin{pmatrix}\sigma_{xx}\\\sigma_{yy}\\\sigma_{xy}\end{pmatrix}=\mathbf{C}_{\boldsymbol{\varepsilon}}\left(\boldsymbol{\varepsilon}_{m}+\zeta\boldsymbol{\varepsilon}_{b}\right), \, \begin{pmatrix}\sigma_{zx}\\\sigma_{zy}\end{pmatrix}=\mathbf{C}_{\boldsymbol{\gamma}}\boldsymbol{\gamma},\\ \sigma_{zz}=C_{z}\varepsilon_{zz}, \, \int_{-1}^{+1} \zeta \mathbf{C}_{\boldsymbol{\varepsilon}}d\zeta=0.
\end{gathered}
\end{equation}
For isotropic materials with the plane-stress condition adopted,
\begin{equation}
\mathbf{C}_{\varepsilon}=\frac{E}{1-\nu^{2}}\begin{pmatrix}1&\nu&0\\\nu&1&0\\0&0&(1-\nu)/2\end{pmatrix}, \, \mathbf{C}_{\boldsymbol{\gamma}}=\frac{5}{6}\frac{E}{2(1+\nu)}\begin{pmatrix}1&0\\0&1\end{pmatrix}, \, C_{z}=E,
\end{equation}
where $E$ is Young's modulus and $\nu$ is Poisson's ratio. With the Jacobain determinant $J$ for global Cartesian coordinates $(X,Y,Z)$ and $(\xi,\eta,\zeta)$ approximated by $J_{o} = J|_{\zeta=0}$, the element strain energy can be written as
\begin{equation}
\begin{aligned}
    \Psi^{e}=\frac{1}{2}\int_{-1}^{+1} \int_{-1}^{+1} \int_{-1}^{+1} &\left[\left(\boldsymbol{\varepsilon}_{m}+\zeta\boldsymbol{\varepsilon}_{b}\right)^{T}\mathbf{C}_{\boldsymbol{\varepsilon}}\left(\boldsymbol{\varepsilon}_{m}+\zeta\boldsymbol{\varepsilon}_{b}\right) \right.\\
    &\left.+\boldsymbol{\gamma}^{T}\mathbf{C}_{\boldsymbol{\gamma}}\boldsymbol{\gamma}+C_{z}\varepsilon_{zz}^{2}\right] J_{o}d\xi d\eta d\zeta.
\end{aligned}
\end{equation}
After integration with respect to $\zeta$:
\begin{equation}
\Psi^{e}=\frac{1}{2}\int_{-1}^{+1} \int_{-1}^{+1} \left(\boldsymbol{\varepsilon}_{m}^{T}\mathbf{C}_{m}\boldsymbol{\varepsilon}_{m}+\boldsymbol{\varepsilon}_{b}^{T}\mathbf{C}_{b}\boldsymbol{\varepsilon}_{b}+\boldsymbol{\gamma}^{T}\mathbf{C}_{S}\boldsymbol{\gamma}+C_{T}\varepsilon_{zz}^{2}\right) J_{o}d\xi d\eta,
\end{equation}
where
\begin{equation}
\left(\mathbf{C}_{m}, \mathbf{C}_{b}, \mathbf{C}_{S}, C_{T}\right)=\int_{-1}^{+1} \left(\mathbf{C}_{\boldsymbol{\varepsilon}}, \zeta^{2}\mathbf{C}_{\boldsymbol{\varepsilon}}, \mathbf{C}_{\boldsymbol{\gamma}}, C_{z}\right)d\zeta .
\end{equation}

\section{Strain Transformation } \label{Strain_Trans}

The Green–Lagrangian strain tensor $\boldsymbol{\varepsilon}$ can be expressed using different basis vectors. In particular,  
\begin{equation}
    \boldsymbol{\varepsilon}=\varepsilon_{i j} \mathbf{e}_i \mathbf{e}_j={\varepsilon}_{m n} {\mathbf{e}}_m {\mathbf{e}}_n,
\end{equation}
where $\mathbf{e}_i$ and $\mathbf{e}_j$ ($i, j = \xi, \eta, \zeta$) are the basis vectors of the natural coordinate frame, and ${\mathbf{e}}_m$ and ${\mathbf{e}}_n$ ($m, n = x, y, z$) are the basis vectors of the local Cartesian coordinate frame. These basis vectors can be computed by: $\mathbf{e}_{\xi}=\mathbf{X}_{o,\xi}, \mathbf{e}_{\eta}=\mathbf{X}_{o,\eta}$ and $\mathbf{e}_{\zeta}=\mathbf{X}_{n}$, as well as 
${\mathbf{e}}_{x} = (\mathbf{X}_{o2}-\mathbf{X}_{o1})/\lVert (\mathbf{X}_{o2}-\mathbf{X}_{o1}) \rVert$, ${\mathbf{e}}_{z} = {\mathbf{e}}_{x} \times (\mathbf{X}_{o4}-\mathbf{X}_{o1})/\Vert (\mathbf{X}_{o4}-\mathbf{X}_{o1}) \Vert / \Vert {\mathbf{e}}_{x} \times (\mathbf{X}_{o4}-\mathbf{X}_{o1})/\Vert (\mathbf{X}_{o4}-\mathbf{X}_{o1}) \Vert$ 
and 
${\mathbf{e}}_{y} = {\mathbf{e}}_{x} \times {\mathbf{e}}_{z}$. 
In this study, the initial state of the origami panels is flat.

The components of the strain tensor in the local Cartesian coordinates can be obtained by 
\begin{equation} 
{\varepsilon}_{m n}={\mathbf{e}}_m\cdot\varepsilon_{i j} \mathbf{e}_i \mathbf{e}_j\cdot{\mathbf{e}}_n=c_{m i} c_{n j} \varepsilon_{i j}
\end{equation} 
where $c_{mi} = {\mathbf{e}}_m \cdot \mathbf{e}_i$ is the cosine of the angle between the local Cartesian basis vector ${\mathbf{e}}_m$ and the natural basis vector $\mathbf{e}_i$.

\newcommand{\norm}[1]{\left\|#1\right\|}
\newcommand{\minn}{\text{min}}
\newcommand{\maxx}{\text{max}}

\begin{algorithm}[H]
\caption{Solving Origami Simulation with Adaptive Increments}
\label{alg:origami_solver}
\begin{algorithmic}[1]

\Require initial\_positions, max\_increments, loaded\_disp

\State $U \gets 0$ \Comment{Gloabl displacement vector}
\State $\lambda \gets 0$ \Comment{Load parameter}
\State $\alpha \gets 1.0$  \Comment{Step size}
\State $\beta \gets 1.0$ \Comment{Relaxation factor}
\State $\gamma \gets 0$ \Comment{Recovery attempts}
\State $\delta \gets 0$ \Comment{Increment counter}
\State $\Gamma \gets 20$ \Comment{Max recovery attempts}

\State $\Delta U_{\text{presc}} \gets$ loaded\_disp $/$ max\_increments \Comment{Prescribed displacement per step}

\While{$\lambda < 1.0$ and $\gamma \leq \Gamma$}
    \State $\delta \gets \delta + 1$, $U_{\text{prev}} \gets U$
    \State $U \gets U + \alpha \cdot \Delta U_{\text{presc}}$
    \State $U[\Phi] \gets 0$ \Comment{Apply Dirichlet BCs at fixed DOFs}
    
    \State $\epsilon \gets \infty$ \Comment{Residual norm}
    \State $\iota \gets 0$ \Comment{Iteration counter}
    \While{$\epsilon >$ tolerance and $\iota <$ max\_iterations}
        \State $K, F_{\text{int}} \gets$ ASSEMBLE\_SOLID\_SHELL\_WITH\_CREASE
        \State $r \gets \lambda \cdot F_{\text{ext}} - F_{\text{int}}$ \Comment{Residual force vector}
        
        \State $\Omega \gets$ find\_free\_dofs()  \Comment{Free dofs}
        
        \State $\Delta U[\Omega] \gets$ SOLVE\_LINEAR\_SYSTEM($K[\Omega, \Omega], r[\Omega]$)
        \State $\Delta U \gets 0$, $\Delta U[\Omega] \gets \Delta U[\Omega]$
        \State $U \gets U + \beta \cdot \Delta U$, $\epsilon \gets \norm{\Delta U[\Omega]}$
        \State $\iota \gets \iota + 1$
    \EndWhile
    
    \If{$\iota \geq ((\alpha > 1) + 1) \cdot \text{max\_iterations} / (\beta + 1)$}
        \State $\gamma \gets \gamma + 1$, $\delta \gets \delta - 1$
        \If{$\gamma \leq 10$}
            \State $\alpha \gets \alpha / 2$
        \Else
            \State $\alpha \gets \max(\alpha, 1) \cdot 1.5$, $\beta \gets \beta \cdot 0.75$
        \EndIf
        \State $U \gets U_{\text{prev}}$
    \Else
        \State $\lambda \gets \lambda + \alpha / \text{max\_increments}$, $\gamma \gets 0$, $\beta \gets 1.0$
        \State $\alpha \gets \begin{cases} 
            \min(\alpha \cdot 1.1, 1) & \text{if } \alpha < 1 \\ 
            \max(\alpha \cdot 0.9, 1) & \text{otherwise}
        \end{cases}$
    \EndIf
\EndWhile

\end{algorithmic}
\end{algorithm}

The membrane strain $\boldsymbol{\varepsilon}_m = \left[\begin{array}{lll}{\varepsilon}_{mxx} &{\varepsilon}_{myy} & {\gamma}_{mxy}\end{array}\right]^{T} $ and bending strain $\boldsymbol{\varepsilon}_b = \left[\begin{array}{lll}{\varepsilon}_{bxx} &{\varepsilon}_{byy} & {\gamma}_{bxy}\end{array}\right]^{T} $ in the local Cartesian coordinates are related to those in the natural coordinate frame by the following transformation
\begin{equation}
    \left\{\begin{array}{c}{\varepsilon}_{mxx} \\{\varepsilon}_{myy} \\ {\gamma}_{mxy}\end{array}\right\}  
    = \mathbf{T}_{\boldsymbol{\varepsilon}}
    \left\{\begin{array} {c}\varepsilon_{m\xi \xi} \\\varepsilon_{m\eta \eta} \\ \gamma_{m\xi \eta}\end{array}\right\},\,
    \left\{\begin{array}{c}{\varepsilon}_{bxx} \\{\varepsilon}_{byy} \\ {\gamma}_{bxy}\end{array}\right\}  
    = \mathbf{T}_{\boldsymbol{\varepsilon}}
    \left\{\begin{array} {c}\varepsilon_{b\xi \xi} \\\varepsilon_{b\eta \eta} \\ \gamma_{b\xi \eta}\end{array}\right\},
\end{equation}
where $\mathbf{T}_{\boldsymbol{\varepsilon}}$ is
\begin{equation}
    \begin{bmatrix}
    c_{ \xi x} c_{ \xi x} & c_{\xi y} c_{\xi y} & c_{\xi x} c_{\xi y} \\
    c_{\eta x} c_{\eta x} & c_{\eta y} c_{\eta y} & c_{\eta x} c_{\eta y} \\
    c_{\xi x} c_{\eta x}+c_{\eta x} c_{\xi x} & c_{\xi y} c_{\eta y}+c_{\eta y} c_{\xi y} & c_{\xi x} c_{\eta y}+c_{\xi y}c_{\eta x}
    \end{bmatrix}^{-1}.
\end{equation}
The transverse shear strain $\boldsymbol{\gamma} = \left[\begin{array}{ll}{\gamma}_{z x} &{\gamma}_{z y} \end{array}\right]^{T} $ in the local Cartesian coordinates can be transformed by
\begin{equation}
\left\{
\begin{array}{c}
\gamma_{z x} \\
\gamma_{z y}
\end{array}
\right\}
=
\mathbf{T}_{\boldsymbol{\gamma}}
\left\{\begin{array}{c}
\gamma_{\zeta \xi} \\
\gamma_{\zeta\eta}
\end{array}\right\},
\end{equation}
where
\begin{equation}
    \mathbf{T}_{s} 
     = \begin{bmatrix}
    c_{\zeta z} c_{\xi x} & c_{\zeta z} c_{\xi y} \\
    c_{\zeta z} c_{\eta x} & c_{\zeta z} c_{\eta y}
    \end{bmatrix}^{-1} .
\end{equation}
Lastly, the thickness strain $\varepsilon_{z z}$ is transformed by
\begin{equation}
    \varepsilon_{z z}= T_{\zeta} \varepsilon_{\zeta \zeta} = (c_{\zeta z}c_{\zeta z})^{-1} \varepsilon_{\zeta \zeta}.
\end{equation}

\section{Nonlinear Solving Procedure} \label{appendixE}
Our computational model is embedded within a damped Newton solver~\cite{wriggers2008nonlinear} with adaptive load increments to handle quasi-static simulations. The overall nonlinear solution process is outlined in Algorithm~\ref{alg:origami_solver}. In this context, $F_{\text{int}}$ and $K$ denote the internal force vector and the Stiffness matrix, respectively, corresponding to the gradient and Hessian of the shell and the crease energies.

\begin{figure}
  \centering
  \includegraphics[width=\linewidth]{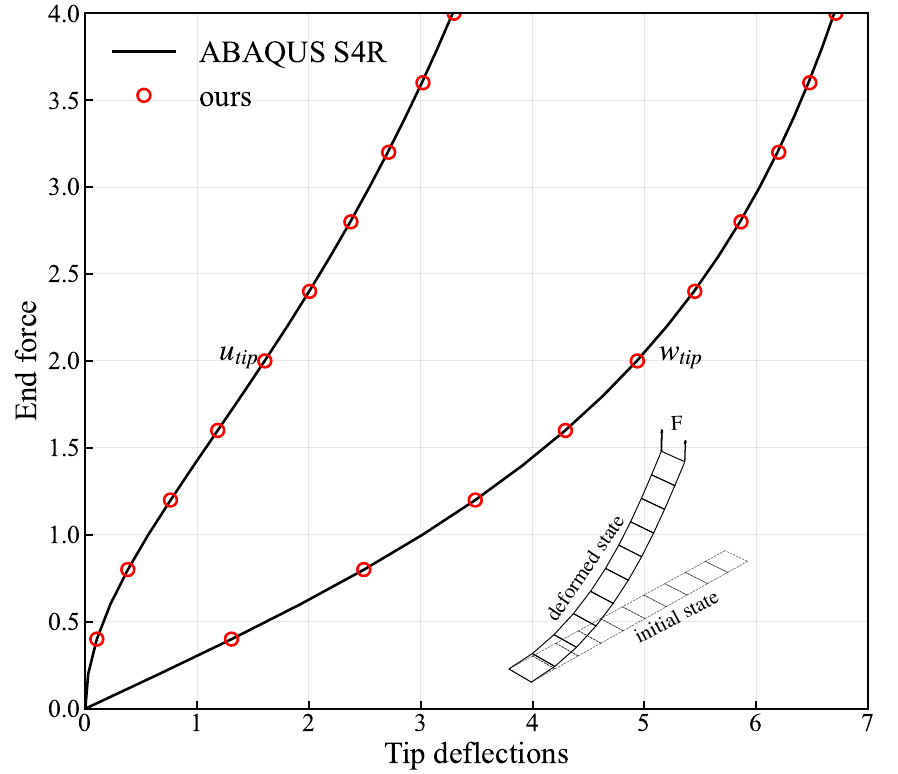}
  \caption{Comparison of the predicted end deflections using the presented solid-shell element with the reference solutions obtained from ABAQUS S4R element. }
  \label{fig:cantilever_deflection}
  \Description{fig:cantilever}
\end{figure}

\section{Supplementary Accuracy Assessment of the Solid-shell Element}

We consider a cantilever beam with dimensions 10 (length), 1 (width), and 0.1 (thickness). The material properties are defined by a Young’s modulus of $1.2 \times 10^9$ and a Poisson’s ratio of 0. The beam is discretized into 10 quadrilateral solid-shell elements. For comparison, a reference solution is obtained using the ABAQUS S4R shell element with a sufficiently fine mesh of $40 \times 4$ elements to ensure convergence.
Figure~\ref{fig:cantilever_deflection} shows the predicted vertical and horizontal tip deflections under a shear force ranging from 0 to $4,000$ ($2,000$ applied at each tip node) in $10$ increments. The results demonstrate a good agreement between our predictions and the reference solution.

\end{document}